# Quantification model of X-ray fluorescence analysis based on iterative Monte Carlo procedure

I. Szalóki*, A. Gerényi, G. Radócz

Institute of Nuclear Techniques, Budapest University of Technology and Economics, Műegyetem rkp. 9, H-1111 Budapest, Hungary,

*Corresponding author: szaloki@reak.bme.hu

**Abstract**

Iterative Monte Carlo algorithm has been constructed and tested for quantification of X-ray fluorescence analysis in order to determine the atomic composition of solid materials. The calculation model uses simulation code MCNP6 that describes the excitation and relaxation phenomenon in the atomic electron shell. The complete analytical procedure was tested by quantitative analysis of standard reference alloy materials. Acceptable agreement was found between the calculated and nominal concentrations within the standard deviations of the concentrations of the major elements. The total duration of the repeatedly performed Monte Carlo numerical computation for the entire analysis was only a few minutes, due to the application of variance reduction procedures available in the Monte Carlo code. A frame software was designed and written in MatLab programming environment for controlling the iterative calculation of the elementary composition of the sample material.

**Introduction**

The energy dispersive X-ray Fluorescence (ED-XRF) analysis is a powerful, quick and non-destructive analytical method for a wide range of materials such as environmental or cultural heritage samples, biological, harmful chemical materials. Part of the analysed samples require specially designed portable XRF devices adapted to the individual physical and chemical properties of the samples. These kind of analytical tasks need low-power X-ray sources and thermoelectric silicon drift detectors (SDD) built in compact structure including the full electronic unit as well [1] that is called as energy dispersive X-ray fluorescence spectrometer (ED-XRF). This compact implementation provides analysis of atomic composition of hazardous and radioactive materials with extensive spatial dimensions. If ED-XRF device is equipped with a CZT based gamma spectrometer it is suitable for complex analysis of radioactive waste materials for both the isotope-selective radioactivity and quantitative atomic composition [2]. Using the mechanical moving part of a 3D printer this type of combined spectrometer was built on the moving frame of a 3D printer with step size of 100 μm in 3D directions. The ED-XRF set-up was constructed in so-called confocal mode [3], where both the X-ray tube and the detector have a limited solid angle with aluminium collimators. The confocal volume can be positioned to any optionally selected surface point of the analysed object. Rather intense X-ray radiation can be achieved with high flux density of synchrotron beam suitable for confocal XRF [4] analysis. In synchrotron based confocal applications, the smallest beam diameter and maximum excitation flux density can be arranged with polycapillary X-ray lenses [5], [6]. The analytical capability such as detection limits, the analyzable volume of laboratory spectrometers based on low-power X-ray tubes is significantly lower than SR-XRF measuring assemblies designed for synchrotron beamlines. In case of samples with irregular spatial shape and heterogeneous atomic composition no standard samples are generally available for empirical calibration of the quantitative composition. The FPM (Fundamental Parameter Method) model is one of the suitable procedures to calculate the elemental composition without standard materials. The FPM is based on physical-mathematical description of the relationship between the detected characteristic X-rays and the concentration of chemical elements. The FPM model considers the energy-dependent absorption of the matrix and internal excitation (enhancement) processes. The FPM is suitable for analysis in a wide range of concentrations ($10^{-4}$-$10^2$ m/m%) within atomic range of 13 < Z. The atomic composition of substances can be determined by Monte Carlo technique as well, simulating the excitation and relaxation processes occurred in the sample and the detector system. One of the earliest model the so called Reverse Monte Carlo (RMC) procedure were developed for X-ray diffraction in order to determine the unknown microstructure [7]. The method is based on variation of space-position of atoms in the crystal structure and calculating the diffraction images than it is compared to the measured one. Depending on this comparative result, and a new MC based approximation is performed again repeating the whole procedure. This MC calculation cycle have to repeatedly carried out while the differences between the empirical pattern and the simulated one become less than limits given preliminarily for a chi-square expression. This RMC principle can be adapted effectively in various type of ED-XRF analysis as well, for example in Electron Probe Micro Analysis (EPMA) of aerosol particles [8], [9]. In this analytical method the excitation of the atoms of the particles are excited by electrons and the cyclically performed RMC approach refers to the concentrations of the chemical elements of the sample. However, the RMC technique can be used efficiently in the Extended X-ray Absorption Fine Structure (EXAFS) studies [10] as well, that is demonstrated by a typical example for investigation of structure of Ni80P20 amorphous alloy. This data evaluation technique fit well to the widely applied ED-XRF technique especially with synchrotron radiation, due to the extremely high X-ray flux of the exciter X-ray beam [11].

**Reverse Monte Carlo algorithm for solution of the ED-XRF quantification problem**

In this study, the Reverse Monte Carlo technique was adapted to approximate the unknown concentrations of a sample material using an appropriate MC model in an iterative cycle. During the recursive process the MC calculation is repeatedly performed continuously while the simulated/measured ratios of the unknown parameters are nearby to the unit within confidence level or the chi-square formula is close to its minimum. For the reverse Monte Carlo (RMC) procedure a validated simulation code is required that describes all the phenomena going on during the excitation and relaxation processes: ionisation, emission of X ray characteristic photons, Auger- and non-radiative processes, scattering effects, energy dependent enhancement and matrix effects. Therefore, the most important requirement against the applied simulation code is that it realistically describes the excitation-relaxation-detection processes. The MC calculation results an ED-XRF spectrum that presumably is within the statistical fluctuation to the measured spectra. Similar behavior is expected also the characteristic intensities of the sample elements. The ultimate goal of RMC-XRF analysis is to determine the relative or absolute concentrations of chemical elements in a sample. The unknown vector variable $\vec{C}$ consists of relative concentrations of the chemical elements in the analysed substance. Let suppose, the sum of weight fractions of the sample elements equal to the unit as it given by (1), which condition meanings that characteristic X-ray radiations are detected for each chemical elements of the sample.

$$\vec{C} \doteq (C_1, C_2, \ldots, C_n) \quad 0 < C_i < 1$$
$$i = 1, 2, \ldots, n \quad \sum_{i=1}^{n} C_i = 1 \quad (1)$$

Define a concentration vector $\vec{C}^{(0)}$, which consists of same chemical element as are in the real sample, but these concentrations are not equal to the real concentrations $\vec{C}^*$ of the sample as it given in the first term of expressions (2). The variable $I_{i,sim}$ and $I_{i,meas}$ are the simulated and measured characteristic X-ray intensities of the $i^{th}$ chemical elements in the sample. Consequently, each $k_i$ does not equal to the unit as it given by second expression in (2).

$$\text{if } \left|C_i^{(0)} - C_i^*\right| \neq 0 \Rightarrow k_i \doteq \frac{I_{i,meas}}{I_{i,sim}\left(\vec{C}^{(0)}\right)} \neq 1$$
$$\Rightarrow C_i^* \neq C_i^{(0)} \frac{I_{i,meas}}{I_{i,sim}\left(\vec{C}^{(0)}\right)} \quad i = 1, 2, \ldots, n \quad (2)$$

Based upon the third inequality formula (2) a new equation can be derived (3) that satisfies the initial conditions of the Banach fixed point theorem [12].

$$\left. \begin{array}{l} I_{i,sim}^{(r)} \doteq I_{i,sim}\left(\vec{C}^{(r)}\right) \\ C_i^{(r+1)} = \dfrac{C_i^{(r)} I_{i,meas}}{I_{i,sim}^{(r)}} \\ \sum_{j=1}^{n} C_j^{(r+1)} = 1 \end{array} \right\} \Rightarrow C_i^{(r+1)} = \dfrac{C_i^{(r)} I_{i,meas}}{I_{i,sim}^{(r)} \sum_{j=1}^{n} \dfrac{C_j^{(r)} I_{j,meas}}{I_{j,sim}^{(r)}}} \quad (3)$$

This recursive formula is the basic expression of an appropriate successive iterative procedure to approach the solution of the RMC problem, where the index *r* is the number of iteration steps. The recursive formula (3) offers a stochastically convergent iterative numerical solution of the RMC problem [8] and it involves the normalisation condition. The number of iteration steps of the recursive numerical calculations are restricted with an optionally given upper limit ($r_{max}$) and parameters $\delta_1$, $\delta_2$ corresponding to relations (4) concern to the concentrations and the characteristic X-ray intensities.

$$\left| C_i^{(r+1)} - C_i^{(r)} \right| < \delta_1 \qquad \left| I_{i,sim}\left(\vec{C}^{(r+1)}\right) - I_{i,meas} \right| < \delta_2$$
$$r = 1, 2, ..., r_{max} \quad (4)$$

First time this general algorithm of RMC quantification model was tested for analysis of aerosol particles [8] with electron probe microanalysis (EPMA) where the simulation code was based on the CASINO [13] that code describes all the physical effects those interact in electron induced X-ray emission processes. Since, the basic principle of RMC algorithm is not system-dependent, perhaps it can also be used in ED-XRF analysis as well.

**Application of MCNP6 code for RMC-XRF**

In order to simulate ED-XRF spectra for analysis of composition of unknown samples the MCNP6 code was used. The main motivation of the selection of MCNP6 for this purpose was the possibility of determination not only the chemical composition of substances, but to identify the isotope selective radioactivity on the basis of gamma radiation of the radioactive isotopes. The MCNP6 is a well-developed and powerful code for stochastic modelling of all the radioactive effects of matters, nuclear reactions and especially detection of gamma radiation. The simultaneous application of MCNP6 for both simulations of atomic excitation and gamma radiation is a great opportunity for the complex and simultaneous investigation of radioactive and harmful materials for determination of both the quantitative composition of inactive part of the sample and the isotope selective radioactivity. The RMC-XRF algorithm has a great advantage if suitable SRM (standard reference material) samples are not available for calibration. The code must be quick enough, namely requires only a few minutes processing time for the full analysis. In practice, it is corresponding to simulate about $10^{10}$ X-ray photons as source events. Unfortunately, for the calculation of this high level of source events the updated computing resource requires about 12-13 hours processing time (non-analogue) that is too long for application of RMC algorithm for the data evaluation in routine XRF analysis. The main problem of the MC calculation is that only a very small portion of the total number of emitted characteristic X-ray photons is detected. In order to decrease the duration of the simulation calculations so called "variance reduction" method has to be used such as DXTRAN SPHERE (DS) [14] method. This deterministic (analogue) algorithm distorts the angles of the atomic emission events in the sample material and the DS code calculates transport-events deterministically. This sub-code modifies the direction of the secondary photons such a way that the majority of the photons propagate into direction of the surrounds of the detector in order to improve the efficiency of the photon-sampling. Due to the application of DS the variance of the weight-factors of the photons calculated by non-analogous MC mode increases, that results the increased variance of the statistics of the total number of detected photons. This numerical effect reduces drastically the efficiency of the simulation calculations that can be neglected by use of the Weight-Window [14] (WW) method. The WW procedure controls the weight-factors of the photons in the phase-space and the result is that the photons contribute to the tally-events having about the same weight-factors. Application simultaneously both (DS and WW) variance-reducing methods in the simulation process the number of required photons could be reduced down to $10^7$. This number decreases the total processing time within 3 order of magnitudes. Thanks to these combined variance-reduction procedures the total calculation time for a single XRF spectrum is no longer than 90 second that offers application of RMC-XRF technique in practice. In order to perform MC simulation to approximate the detected X-ray spectra it is necessary to get information of the energy dependent flux density distribution of the primary (exciter) X-ray beam emitted by the X-ray tube. This type of function was determined empirically [15] in energy range between 5 – 40 keV plotted on Fig. 1.

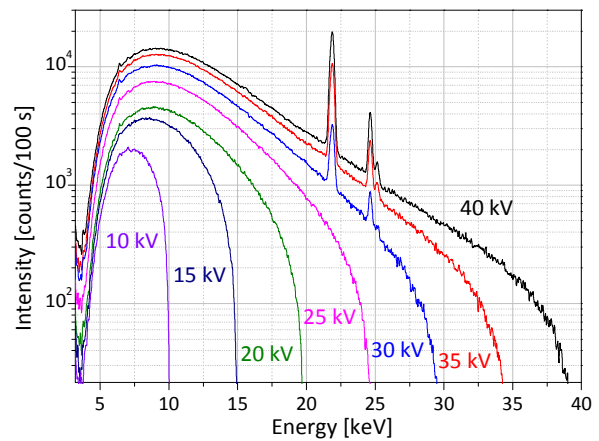

Figure 1. Primary spectra of Amptek Mini-X X-ray tube with Al collimator (length 15 mm, inner diameter 2 mm).

**Convergence of the RMC-XRF calculations**

For investigation of convergence of the numerical RMC calculations RC-34/1 SRM alloy sample was analysed with generating $10^6$ photons in each iteration step and the whole analytical calculation was repeated 10-times. The results of these RMC-XRF numerical calculations are showed by the Table 1. For the comparability of the calculated concentrations the Table 1 exhibit the measured and calculated values determined by RMC-XRF and Spark Optical Emission Spectrometry (SOES). The standard deviations of the RMC-XRF calculations were found in range of 0.01-0.3 m/m% between 0.5-52.0 m/m% of concentrations, while the average difference is between RMC-XRF and SOES was 1.15 m/m%. Errors of the SOES concentrations were not available for the SRM samples.

Table 1. Results of RMC-XRF model calculations for SRM RC-34/1. The iterative RMC calculations were repeated 10-times.

| Elements | Concentrations [m/m%] | | | |
|---|---|---|---|---|
| | SOES | RMC mean | RMC SD | \|RMC-SOES\| |
| Cr | 16.3 | 19.31 | 0.19 | 3.01 |
| Mn | 7.95 | 8.34 | 0.21 | 0.39 |
| Fe | 51.5 | 50.12 | 0.27 | 1.38 |
| Co | 0.32 | 0.69 | 0.05 | 0.37 |
| Ni | 20.8 | 18.62 | 0.22 | 2.18 |
| Cu | 1.93 | 1.89 | 0.12 | 0.04 |
| Nb | 0.44 | 0.41 | 0.01 | 0.03 |
| Mean of differences | | | | 1.15 |

The key issue of the numerical process of RMC-XRF: is there convergence calculation algorithm, and if so, what speed of computation can be achieved? Namely, how much number of successive approximation steps is necessary for approaching the theoretical solution within an optionally selected upper and lower boundaries? In order to investigate this property of the successive approximation the quantitative atomic composition of the SRM RC-34/1 was calculated. In the first test the initial concentrations of each chemical element were set as 16.67 m/m%. The concentrations converged very quickly and reached the optionally selected criterion after the 4-7 iteration steps. After the 4-5th iteration steps only the statistical uncertainty, that is originated from the stochastically approach, influences the calculated concentrations (Fig. 2) and the algorithm approximates the SOES concentrations within the confidence interval of 0.95 significance at Mn, Cu and Nb elements.

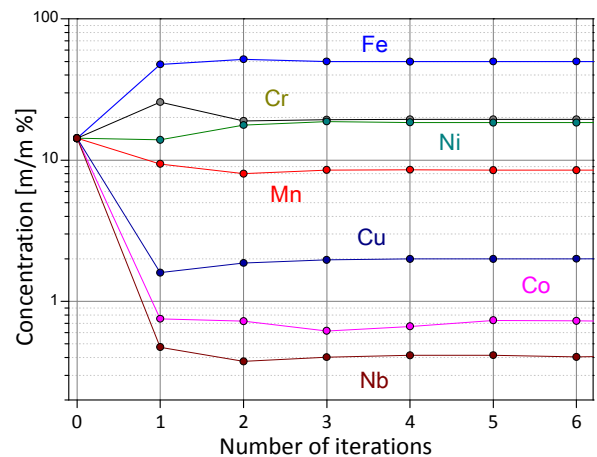

Figure 2. Convergence of the RMC-XRF numerical calculation algorithm on SRM RC-34/1 sample. The initial concentrations were set 16.67 m/m% for each chemical elements.

**Evaluation of the simulated XRF spectra**

The simulated energy dispersive X-ray spectra were calculated by the MCNP6 transport software that acts as the core code in the RMC algorithm. During the iterative calculations the peak broadening effects caused by electronic noise and the signal shaping in the spectrometer electronic units can be considered by so called GEB parameter in the simulation processes [14]. However, the evaluation of the simulated spectra using dedicated software's such as WinQxas [16], bAxil [17] or pyMCA [11] in every iteration step increases the necessary processing time and it requires the intervention of the operator person as well. The background correction of the intensities of the simulated characteristic peaks can be calculated by the differences of the average values of the neighbour channels (before the application of the GEB routine). The main benefit of this background correction mode is the lack of diffraction peaks in the spectra that do not disturb the evaluation procedure of the simulated

X-ray spectra. At the end of the iteration procedure in the simulation optionally can be used the GEB parameter that produces spectrum with broadening effects for comparability of measured and simulated versions (see Fig. 3).

In ED-XRF spectrometry the detector window is usually made from Be or thin polymer layer and the sample-window volume can be filled by air/He or it is evacuated. These conditions determine the lower limit of the atomic number because under this value the characteristic radiations of chemical elements cannot be detected. This fraction of the sample matrix constitutes low-Z elements that is called as dark matrix

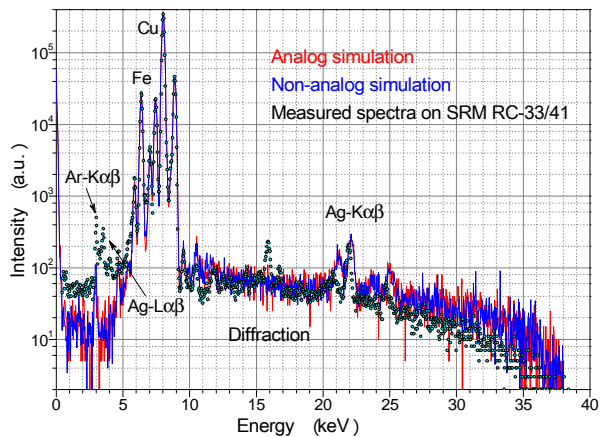

Figure 3. Results of analogue (13 hours), non-analogue (90 s) simulated and measured X-ray spectra of RC-33/41 alloy SRM.

[11], [18], [19]. Due to the absorption effect of the dark matrix on the characteristic radiations of the detectable elements has to be taken account during MC calculations.

In order to perform the RMC algorithm i.e. the numerical calculations and control of the full simulation approximation with MCNP6 software package a frame code was constructed in MatLab programing environment. If the concentrations of some chemical elements of the sample are known preliminary it can define these values as a constant parameter during the iteration calculations. In each iteration step the code visualises the current calculated data: concentrations, number of iteration steps, chi-square and the difference of the last two concentrations for every element comparing to the preliminary given limits. For calculation of the measured characteristic intensities of the sample elements the spectrum had to be evaluated by the dedicated WinQxas [20] software.

**Validation of the XRF-RMC algorithm and code**

For validation of the RMC-XRF technique a set of SRM metal alloy samples were used [21] (Worldwide Analytical Systems AG, Wellesweg 31, D-47589 Uedem, Germany). The excitation of the sample elements was carried out by an Amptek Mini-X-Ag low-power (4 W) transmission type X-ray tube [2]. The maximum of the high voltage was 40 kV and the Ag anode was an ideal selection for K-lines excitation of the chemical elements in the range of atomic number 13-50 (except of Ag), and using L-lines elements over 51. The Be window of the Canberra silicon drift (SD) detector was 25 µm and for each sample the measuring time was set 500 s, the accelerator high voltage was set as 40 kV and the anode current 5 µA. In Fig. 3 both the measured and simulated XRF spectra of the RC-33/41 sample were plotted for comparison. The Ar-K$\alpha\beta$ lines and some diffracted peaks are missing in the simulated spectra due to the lack of these effects in the simulated model, but in the measured version both of them can be recognised. The diffraction peaks should cause evaluation problem if they overlap with characteristic X-ray peaks of any chemical elements in the sample. In this case the diffracted peaks can be consider during the fitting procedure of the measured spectra that these peaks are substituted by a virtual chemical elements giving the exact energy of this peak.

Table 2. Concentrations of elements in SRM alloy samples determined by RMC-XRF and SOES.

| Elements | Concentrations (m/m%) | | | |
|---|---|---|---|---|
| | RC-38/20 | | RC-33/41 | |
| | SOES | RMC | SOES | RMC |
| Mg | 0.0136 | 0.0136 | < 0.01 | ---- |
| Al | < 0.01 | ---- | 10.5 | 10.5 |
| Si | < 0.01 | ---- | 0.226 | 0.226 |
| Cr | < 0.01 | ---- | 0.0116 | ---- |
| Mn | 0.728 | 0.64 | 0.304 | 0.24 |
| Fe | 0.852 | 1.14 | 4.05 | 4.91 |
| Ni | 31.3 | 32.79 | 4.64 | 5.48 |
| Cu | 67.0 | 65.44 | 79.7 | 78.94 |
| Zn | 0.015 | ---- | 0.371 | 0.735 |

In Table 2 shows the result of RMC quantification comparing to the nominal concentrations. The simulations were carried out with generating $10^6$ photons in each iteration steps. In the tables blue colour assigned elements whose characteristic photons could not be detected due to the very low energy (E<1.5 keV) therefore they were used as elements (C, O, N, Be, Na, Mg) with known concentrations those were constant during the iteration process. At the case of Pb and Bi the K-lines cannot excite with the applied set-up (40 kV accelerator high-voltage) only the L- and M-lines can be excited. Unfortunately, in this moment the 6th version of the MCNP software package the DXTRAN variance reduction procedure can not apply for the simulation of L-lines. In Table 2 the symbol "<" indicates that the concentration is less than the minimum detection limit (MDL) [22] of the analytical method. The larger deviations can be recognised at the minor elements (C < 1 m/m%) due to the significantly larger statistical uncertainty of the low intensity peaks. The effect is more influence the MDL values when the X-ray characteristic peaks of major and minor (or trace) elements are overlap that results the larger error of low-intense peaks. This does not a major issue since this equipment was designed and built for quick, non-destructive and quantitative XRF analysis of radioactive materials and samples collected by safeguards investigations. This aim primarily requires the measurement of the main components of the sample that type of analytical method do not requires any chemical or destructive preparation of the investigated objects.

**Optimization of confocal set-up by MC calculation**

In XRF analytical technique, the confocal measuring setup is based on the limitation of the divergence of the exciter (primary) and the secondary X-ray beams using collimators, mono-, or polycapillary X-ray lenses. If the axes of the primary and the secondary beams are crossing each other (CP = confocal point) the detector will sense only those photons which are emitted from a limited volume (CV = confocal volume) located around CP and most part of the photons emitted other part (that CV) of the sample volume cannot [4]. The size of the CV depends on the diameters of the two beams, which parameter depends on the optical properties of the X-ray lenses or collimators used in the confocal setup. The principle of a simple confocal setup and its experimental implementation can be seen on Fig. 4. that is built in the 3D-XRF device. The collimators for both X-ray beams were made from pure (99.99 m/m%) aluminium tubes, those can be inserted tightly into each other providing the required beam diameter. The collimator system of the SDD has a special inner structure with variable geometry in order to decrease the flux of the scattered radiation to improve signal-to-noise ratio of the detected X-rays. Applying sequentially Al tubes with combination of different inner diameter optionally, an arbitrary CV can be formed. In order to simplify and accelerate the MC calculations only those X-ray beam-paths were considered whose travels through the collimator without scattering on the collimator wall. Corresponding to our preliminary expectations that simulation which carried out with conical-type collimator provides more intense beam than it was experienced using constant diameter of the collimator tubes. The length of the constant-diameter collimators of the X-ray tube was set 50 mm (see Fig. 4). The inner diameters of the additionally matched collimator items were 0.5, 1.1 and 2.0 mm. The entering diameter of the conical-type collimators was 4 mm, while at the output end of this parameter was set as 0.5, 1.1 and 2.0 mm. At the case of the conical inner shape of the collimator the output flux is higher than at straight collimator with constant diameter. - The result of simulation for the intensity distribution of the primary X-ray beam is plotted on Fig. 5. On the basis of these result of the MC calculations, it can be concluded that the conical collimators provide higher flux, than straight versions. The reference point for the MC calculations was set on the collimator's symmetry axis 10 mm from the end of the collimator. From the results described above, it appears that both the intensity of the beams and the FWHM of the primary X-ray beams are strongly dependent on the geometric design of the measuring set-up. Therefore, the geometrical parameters (collimators, distances etc.) influence significantly the lateral resolution of the 3D-XRF [2] system. An optimal collimator design can be defined that maximizes the flux of the excitation beam for a given beam size, that minimizes the MDLs (method detection limit) that corresponds to a certain measuring set-up and parameters. The largest X-ray flux and smallest FWHM of the beam were found at the case of conical collimators. This type of collimators was emulated by set of straight capillaries with different inner diameters. The experimental and simulated results were found to be in good agreement: diameter and flux density of the collimated X-ray beams that is showed by Fig. 8. The inner diameter of the capillaries was 0.5, 1.1 and 2.0 mm and they were fit tightly into each other. Based upon the simulation results this stepped collimator system, suitable for maximizing the output flux and simultaneously keeping the beam FWHM at a minimum value. For the 50 mm length of the external copper alloy collimator, the optimum arrangement was found between 5 and 10 mm for a tube with a diameter of 1.1 mm, and between 2.5 and 5 mm for a tube with an inside diameter of 0.5 mm. The profile of the output beam is saturated at radius 200-300 µm while the FWHM is between 500 and 700 µm. Due to the confocal measuring set-up, the exact volume and dimensions according to the three coordinates of the irradiated volume in the sample (confocal volume) is an essential parameter for this 3D technique, therefore this was determined empirically and calculated by simulation. Experiment was performed on a pure Cu disc with thickness of 25 µm and it was scanned with 100 µm step size along x, y and z axis of the 3D-XRF device [2].

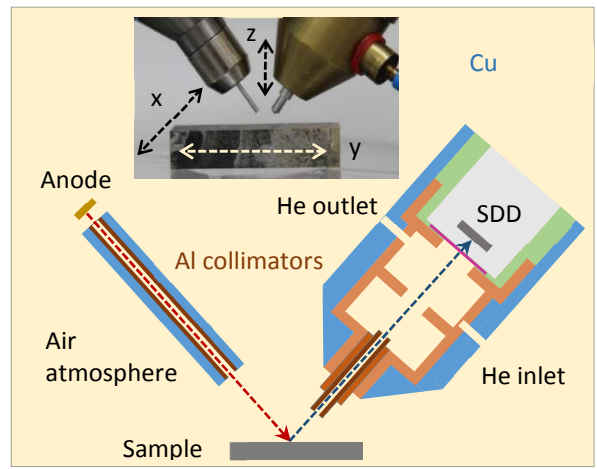

Figure 4. The experimental implementation of confocal set-up in the 3D-XRF device and the input model for MCNP6 simulation code.

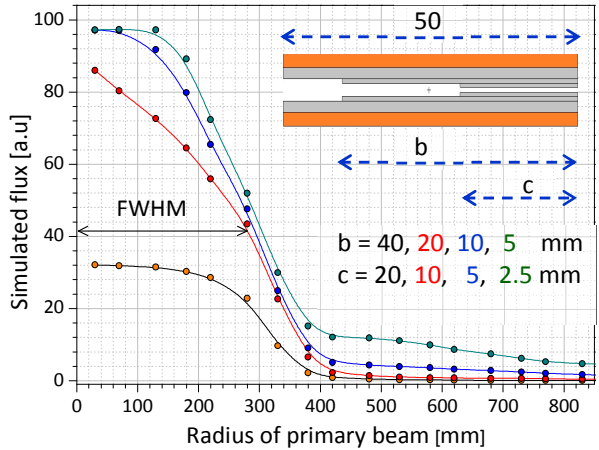

Figure 5. Flux-distribution of the excitation X-ray beam at the output of the collimator mounted in front of the X-ray tube.

Corresponding to the XRF set-up the x and y scanning resulted an increasing behaviour of the Cu-Kα intensity that saturated, those curves can be fitted by an Erf function, while the z-scan curve was fitted by Gauss-function. These numerical results indicated that the profile of the X-ray beam can be described mathematically in all of the tree direction by Gauss-shape. The graphics results of the experiments and the simulation are plotted on Fig. 10 for the x and z space coordinates. The inner diameter of the applied straight collimator was set 1.1 mm. The FWHM of the detected area of the irradiated spot on sample surface were FWHMx=732 µm and FWHMz=890 µm. The FWHMs were determined by the Cu-Kα detected intensity distribution.

**Conclusions**

New iterative MC procedure has been constructed and tested for quantitative evaluation of energy dispersive X-ray fluorescence data for determination of elementary composition of solid materials. The core of the calculation model is based on MCNP6 simulation code describing the excitation and relaxation processes occur in the atomic electron shell. In order to solve the reverse MC model a numerical iterative algorithm was developed based on the successive approximation technique and the Banach fixed point theorem. For the control of the calculation process a frame software was constructed and written in MatLab system. The biggest problem in making the MCNP code suitable for the analytical task was the duration of the numerical calculations. Initially, the simulation time in each iterative step reached 10-12 hours. The significant acceleration of the computing process was achieved by application of the so-called variance reduction techniques. As a result, the computation time of one iterative step was reduced to 1-2 minutes. The full RMC-XRF analytical method was tested by analysis of standard reference alloys. Acceptable agreement was found between two dataset of concentrations determined by RMC-XRF method and SOES analytical technique. The differences between the two set of concentrations were found within the standard deviations of the relative concentrations of the major elements. This RMC-XRF algorithm and complete analytical method is suitable for the practical application for daily routine quantitative analysis performed by simple laboratory XRF devices.

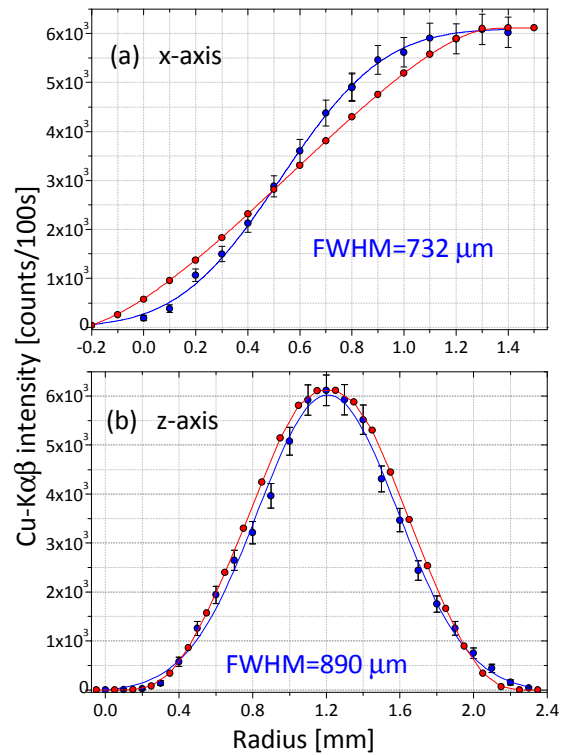

Figure 6. Measured and simulated Cu-Kα intensities along the x and z axes. The inner diameter of the collimators was 1.1 mm. Red curves are the simulated and the blue one is the measured data.

**Acknowledgments**

This work has been partly carried out in the frame of VKSZ-14-1-2015-0021 Hungarian project supported by the National Research, Development and Innovation Fund and by Hungarian Atomic Energy Authority.